\documentclass[%
 aip,
 amsmath,amssymb,
 reprint,%
]{revtex4-1}

\usepackage{graphicx}
\usepackage{dcolumn}
\usepackage{bm}

\usepackage[utf8]{inputenc}
\usepackage[T1]{fontenc}
\usepackage{mathptmx}
\usepackage{etoolbox}
\usepackage{bigints}
\usepackage[dvipsnames]{xcolor}
\usepackage{comment} 
\usepackage{multirow}

\makeatletter
\def\@email#1#2{%
 \endgroup
 \patchcmd{\titleblock@produce}
  {\frontmatter@RRAPformat}
  {\frontmatter@RRAPformat{\produce@RRAP{*#1\href{mailto:#2}{#2}}}\frontmatter@RRAPformat}
  {}{}
}%
\makeatother
\begin{document}


\title{Solubility and dissociation of ionic liquids in epoxides and cyclic carbonate by molecular dynamics simulation}

\author{S. Dorado-Alfaro}
\affiliation{Departamento de Ingeniería Química, Universidad Autónoma de Madrid, Calle Francisco Tomás y Valiente, 28049 Madrid, Spain}

\author{E. Hernández}
\affiliation{Departamento de Ingeniería Química, Universidad Autónoma de Madrid, Calle Francisco Tomás y Valiente, 28049 Madrid, Spain}

\author{J. Algaba}
\affiliation{Laboratorio de Simulaci\'on Molecular y Qu\'imica Computacional, CIQSO-Centro de Investigaci\'on en Qu\'imica Sostenible and Departamento de Ciencias Integradas, Universidad de Huelva, 21006 Huelva Spain}

\author{P. Navarro}
\affiliation{Departamento de Ingeniería Química, Universidad Autónoma de Madrid, Calle Francisco Tomás y Valiente, 28049 Madrid, Spain}

\author{F. J. Blas}
\affiliation{Laboratorio de Simulaci\'on Molecular y Qu\'imica Computacional, CIQSO-Centro de Investigaci\'on en Qu\'imica Sostenible and Departamento de Ciencias Integradas, Universidad de Huelva, 21006 Huelva Spain}

\author{José Palomar$^{*}$}
\affiliation{Departamento de Ingeniería Química, Universidad Autónoma de Madrid, Calle Francisco Tomás y Valiente, 28049 Madrid, Spain}

\begin{abstract}
Climate emergency has led to the investigation of CO$_{2}$ valorization routes. A competitive process included in this framework is the catalytic CO$_{2}$ cycloaddition to epoxides, to produce cyclic carbonates. Halide-based Ionic liquids (ILs) have been postulated to be a competitive choice. Nevertheless, the structure-performance relation for different ILs is still a topic of debate, being the cation-anion dissociation constant a key descriptor. In this work, the ions effect is tackled by Molecular Dynamics (MD) simulations. Propylene oxide and carbonate force fields were tested and used for 1,2-epoxyhexane and hexylene carbonate force field construction, while ILs were modelled by the CL\&P force field. Solubilities in an epoxide-carbonate medium were tested for ILs composed of [$\mathrm{N4444^+}$] or [$\mathrm{N2222^+}$] cations combined with the halide anions: Iodide [$\mathrm{I^-}$], Bromide [$\mathrm{Br^-}$] and Chloride [$\mathrm{Cl^-}$]. Results showed that [$\mathrm{N2222^+}$] cation-based ionic liquids were insoluble in the epoxide/carbonate medium, whereas [$\mathrm{N4444^+}$] cation-based ionic liquids demonstrated diffusion. Reaction medium interactions were studied between key atoms for experimentally soluble ILs. It was found that cation-anion interaction follows the catalytic activity trend, being [$\mathrm{I^-}$] the halide anion less associated with ([$\mathrm{N4444^+}$], [$\mathrm{bmim^+}$] and [$\mathrm{emim^+}$]) cations. A correlation between the first peak integration of the radial distribution functions and the experimental yields (including [$\mathrm{N1111^+}$][$\mathrm{Br^-}$]) could be established with a regression coefficient of 0.86. Additionally, [$\mathrm{I^-}$] based ILs displayed a better interaction between the cation and the epoxide oxygen, phenomena linked to epoxide activation and intermediates stabilization. Therefore, the path towards the understanding of this catalytic system has been widened.
\end{abstract}

\maketitle
$^*$Corresponding author: pepe.palomar@uam.es






\section{Introduction}
\label{sec:sample1}
Since the onset of the Industrial Revolution, human activities have contributed significantly to greenhouse gas emissions. This steady increase in global concentrations of carbon dioxide (CO$_{2}$) and methane, which is made patent by global reports \cite{lee2023ipcc}, has been strongly correlated with environmental changes such as the global temperature rise. These changes not only damage terrestrial, ocean, and freshwater ecosystems, but also significantly impact human societies through the intensification of natural disasters, decline in mental health, and malnutrition, as stated by IPCC reports \cite{noaa2021department,lee2023ipcc}. Social concern about climate change has prompted developed countries to formulate global commitment plans that limit global warming to 1.5ºC by gas emissions reductions, aiming for net-zero by 2050 \cite{netzero2050}. This goal would be achieved by the development of clean energy sources as a long-term solution, coupled with the technological development and operation of Carbon Capture and Utilization (CCU) and Carbon Capture and Storage (CCS) plants. CCUs are present in new energetic carriers development like biogas, which needs an upgrading process for quality standards compliance, on-site carbon capture, and carbon capture delocalization, which can result in net-zero or negative carbon emissions by performing Direct Air Capture (DAC) away from the carbon source. Thus, there is a need to convert CO$_{2}$ into a resource, a process in which CCU plants become essential as a mitigation technique in the environmental transition framework towards more sustainable societies \cite{netzero2050, TrackClean, CCUSProj}.  \\

Obtaining a valuable product from CO$_{2}$ is therefore crucial to achieving the economic success of CCU processes, being cyclic carbonates and polycarbonates a strong and frequently investigated approach in the literature \cite{marcolongo2021stepping, pescarmona2012challenges, kamphuis2019co}. These compounds can be synthesized by the CO$_{2}$ cycloaddition to epoxides, in a reaction that achieves 100\% atom efficiency \cite{kamphuis2019co, sakakura2007transformation}. Among the most important applications of cyclic carbonates are their use as electrolytes for lithium-based batteries, in paint stripping, coatings, and lubricants, along in the synthesis of derived chemicals like polyurethane \cite{su2019cyclic, pescarmona2021cyclic, north2010synthesis, yu2021aliphatic}. Cyclic carbonates can also be used as green solvents in membrane separation or simply in separation stages in classical hydrocarbon separations and pharmaceutical extractions from water as useful hydrophobic solvents \cite{rasool2019applicability, belinchon2022biocarbonates}. Lastly, cyclic carbonates can be used for polycarbonate production by means of the Asahi Kasei non-phosgene process with flexibility on the incorporated chemical groups \cite{kamphuis2019co}. Therefore, the CO$_{2}$ cycloaddition to epoxides constitutes a promising route for CO$_{2}$ valorization whose optimization and challenge addressing are mandatory to achieve competitiveness with the less sustainable conventional routes capable of synthesizing the same products, which has led to rigorous process design and optimization research for different cyclic carbonates and conditions \cite{hernandez2022universal, belinchon2022reaction, belinchon2024extending, palomar2024process}.

An important aspect of the CO$_{2}$ cycloaddition to epoxides is the need for homogeneous or heterogeneous catalysts \cite{pescarmona2021cyclic}. Cyclic carbonates are usually obtained by using Lewis bases without a Lewis acid, such as halide salts or halide-based Ionic Liquids (ILs), or by combining Lewis acids with bases without the presence of metals \cite{kamphuis2019co,alves2017organocatalyzed, alassmy2019role}. Metal-free catalysts such as ILs are promising due to their negligible vapor pressure, adequate thermal stability, and very importantly, they offer a vast tunability by cation and anion change and functionalization, enabling chemical structure design for specific purposes and applications as the aforementioned catalytic step \cite{moya2018co2, palomar2024process, zeng2017ionic}. Additionally, ILs can offer both physical and chemical CO$_{2}$ capture, enabling their applicability at the lowest CO$_{2}$ partial pressures conditions through the so-called integrated CCU concept \cite{zanatta2023direct, dorado2024exploiting}. All these advantages have led to their characterization by both experimental and computational techniques, in which different cation and anion combinations have been tried. The most common cations correspond to imidazolium-, ammonium-, phosphonium-, pyridinium, and bis(triphenylphosphine)iminium-based groups \cite{kamphuis2019co, alassmy2019role, chen2019conversion} the first two being the groups studied in this work. In terms of the anion, halides are the most extended, especially [$\mathrm{Cl^-}$], [$\mathrm{Br^-}$] and [$\mathrm{I^-}$]. \cite{chen2019conversion, li2023machine, hernandez2024dissociation}.  \\

\begin{figure}[h]
    \centering
    \includegraphics[width=\linewidth]{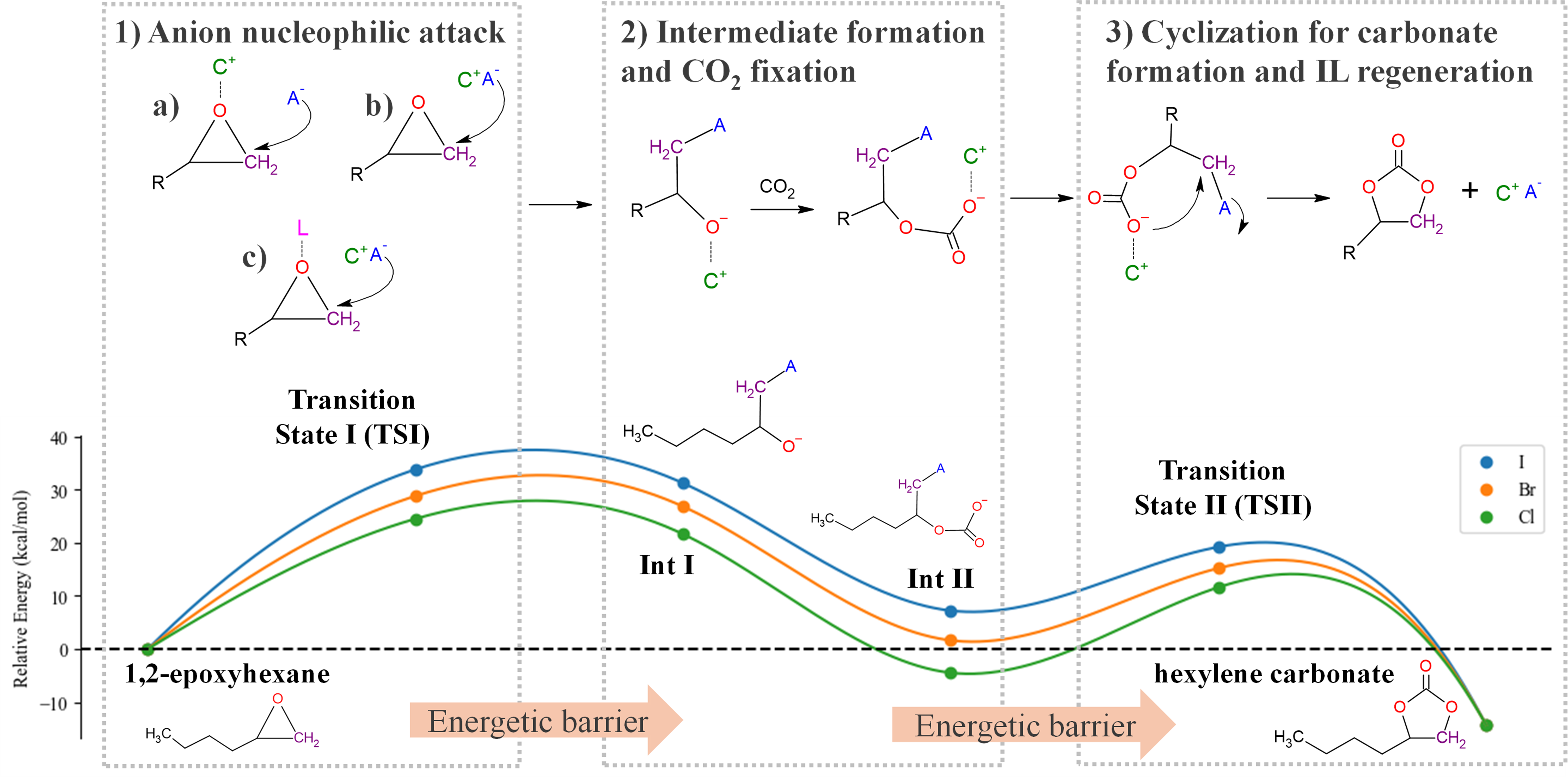}
    \caption{Most extended reaction mechanism for CO$_{2}$ cycloaddition to epoxides by Ionic Liquids catalysts. Adapted from \cite{hernandez2024dissociation} }
    \label{fig:mecanismo}
\end{figure}

Several reaction mechanisms have been attributed to CO$_{2}$ cycloaddition to epoxides by halide-based IL catalysts. The most accepted one is depicted in Fig.\ref{fig:mecanismo}. This mechanism starts with the opening of the epoxide ring by the nucleophilic attack of the halide anion, forming an intermediate species (Int I) in which the anion is covalently bonded to the epoxide. This step is still a topic of controversy due to the fickle behavior of the catalytic performance displayed by the cation selection. It has been proposed that the cation, due to its acidic nature, first interacts with the oxygen of the epoxide by hydrogen bonding, thereby activating the epoxide for reaction and empowering the nucleophilic attack, playing the role of an intermediate stabilizer afterward \cite{bobbink2018intricacies, sun2007density, cheng2015effect, anthofer2014cycloaddition, yan2021fixation, clark2023solvation}. This hypothesis is supported by the experimental catalytic activity improvement displayed by the presence of hydrogen bond donor solvents \cite{wang2012experimental, hernandez2022universal}. Possibilities about the cation influence are (see Fig. \ref{fig:mecanismo}): a) The cation activates the epoxide first by interaction with the oxygen. b) The nucleophilic attack of the anion is independent of the cation. c) A coordination species (L) activates the epoxide \cite{clark2023solvation} before the anion attack (presumably another cation). In the next step, Int I captures CO$_{2}$ and adds it to the chain, displacing the negative charge and forming a new intermediate (Int II). Finally, intramolecular cyclization occurs, forming the cyclic carbonate and regenerating the catalyst by expulsing the anion back into the reaction medium \cite{kamphuis2019co, pescarmona2021cyclic, alves2017organocatalyzed, hernandez2024dissociation, bobbink2018intricacies, butera2023carbon, yan2021fixation, xia2020theoretical, della2016glycidol, carrillo2015molecular, clark2023solvation}.  \\

The experimental catalytic activity of halide-based ILs has been a topic of great discussion in the literature, being the absence of systematic and comparable experiments, until the latest studies\cite{hernandez2024dissociation}, one of the reasons for the lack of understanding of the ionic effect. The most common experimental trend in terms of the anion finds that the catalytic performance follows the order [$\mathrm{I^-}$] $>$ [$\mathrm{Br^-}$] $>$ [$\mathrm{Cl^-}$] \cite{bobbink2018intricacies, alves2017organocatalyzed, alassmy2019role, hernandez2024dissociation, sun2009water}. This order is the opposite of the one expected from DFT calculations, which finds that intermediates with [$\mathrm{I^-}$] are more energetic than those with [$\mathrm{Br^-}$] and [$\mathrm{Cl^-}$] for 1,2-epoxyhexane reactant, as shown in Fig.\ref{fig:mecanismo}. On the other hand, several studies have found the opposite trend even for benchmark fixed cations like [$\mathrm{N4444^+}$], or situations in which different cations and epoxides (steric hindrance might play a significant role between epoxides) result in [$\mathrm{Br^-}$] being the best anion. \cite{bobbink2018intricacies, ju2008catalytic}. In terms of the cations, their versatility has led to very different interpretations, yet without an agreement on how they affect the catalytic activity. It is known that the cation severely affects the catalytic activity, given that by keeping the anion fixed while trying for different cations, the epoxide conversion varies in almost the whole possible spectrum \cite{hernandez2024dissociation}. Generally, improved catalytic activities have been reported for bulkier cations with longer side chains \cite{ju2008catalytic, alassmy2019role, lee2008synthesis}, even though this trend sometimes reverses \cite{xiao2014protic} or 
 remains undecided \cite{bobbink2018intricacies, ju2008catalytic}. Additionally, different cation and anion combinations remarkably affect the ionic liquid solubility in the epoxide/carbonate medium, with a severe impact on the resulting catalytic activity \cite{hernandez2024dissociation}.  \\

In the search for useful descriptors capable of predicting the experimental catalytic activity of halide-based ionic liquids, computational studies with different approaches have been carried out. DFT calculations for [$\mathrm{bmim^+}$][$\mathrm{Br^-}$] showed that epoxide ring opening in propylene oxide (PO) was the rate-determining step \cite{xia2020theoretical}. Additionally, epoxide orientation against the catalyst was found to be significant for the case of styrene oxide \cite{marmitt2015dft}, showing that the epoxide substituents can also affect energetically. More nucleophilic anions displayed lower energetic barriers, which were further lowered by the cations ability to form hydrogen bonds with the epoxide \cite{xia2020theoretical, sun2007density, butera2023carbon, sun2020protic}. Following the same cation principle, water incorporation as solvent resulted in a reduction of the energetic barriers \cite{xia2020theoretical}. Nevertheless, contradictions between computational studies and experimental results are very notable. For example, an insensitivity to the alkyl side chains length \cite{sun2007density} or cation hydrogen bonding not playing a key role in epoxide ring opening \cite{bobbink2018intricacies}, which aligns with the poorer performance of cations containing hydroxyl groups \cite{hernandez2021understanding}. Therefore, a balance between the nucleophilic character of the anion (in charge of epoxide ring opening) and the cation-anion interaction (also favoured) is expected to be best \cite{bobbink2018intricacies}. The discrepancies found between DFT calculations and experimental results are probably due to the inability of DFT calculations to consider bulk properties such as solubility, solvation, and transport phenomena. Alternative studies based on DFT/COSMO-RS approaches were successfully used for cation-anion interaction description through the IL dissociation equilibrium constant (K$_{D}$) calculation \cite{hernandez2024dissociation, lee2015priori}, which has been experimentally validated by the dissociation degree ($\xi$), obtained from experimental measurements of conductivity ($\lambda$) in the epoxide and dissolved IL catalysts \cite{hernandez2024dissociation}. The K$_{D}$ parameter corresponds to the equilibrium constant between associated and dissociated ions as the reaction \( \text{CA} \leftrightarrow \text{C}^+ + \text{A}^- \), in which C, A, and CA stand for the Cation, the Anion, and the Cation-Anion associated pair respectively. Following this definition, K$_{D}$ could be calculated theoretically. K$_{D}$ is to date the most successful catalytic activity descriptor, as high regression coefficients between carbonate yield and K$_{D}$ have been established \cite{hernandez2024dissociation}. Reported K$_{D}$ values and experimental conductivity results are lower than expected, which conducts to the thought of ions as associated species, surely affecting the reaction mechanism interpretation. All these findings reveal the important interplay between all the molecular interactions present in this catalytic system, giving a special emphasis to the cation-anion interaction. Alternative simulation methods capable of accounting for these effects such as classical Molecular Dynamics (MD) can therefore be thought to be enlightening towards the understanding of the ions influence on the experimental catalytic activity. Literature concerning to MD simulations of these systems has been able to link experimental catalytic performance with CO$_{2}$ availability around the epoxide ring \cite{carrillo2015molecular}. Additionally, cation cooperation for ring opening has also been proposed as a result of the detailed analysis of Radial Distribution Functions (RDFs) between the simulated species \cite{clark2023solvation}.  \\

The CO$_{2}$ cycloaddition to epoxides catalyzed by halide-based ILs is, therefore, a complex reactive system in which multiple inter-playing phenomena define the resulting catalytic activity and cyclic carbonate yield. In this work, MD simulations for ILs composed of [$\mathrm{emim^+}$], [$\mathrm{bmim^+}$], [$\mathrm{N1111^+}$], [$\mathrm{N2222^+}$], and [$\mathrm{N4444^+}$] cations and the three halide anions [$\mathrm{Cl^-}$], [$\mathrm{Br^-}$], and [$\mathrm{I^-}$] are performed in the search of chemical descriptors capable of explaining the experimental trends directly from simulation. The different ionic liquids resulting from the previous cation and anion matrix selection cover both experimentally soluble and insoluble ILs in an epoxide and carbonate medium. Additionally, the experimental catalytic activities measured by the experimental yields range from 0\% to 91\% for the variety of selected ILs. Simulation results are then compared with the most systematic available experimental results \cite{hernandez2024dissociation}. The solubility in 1,2-epoxyhexane (H-Epox) and hexylene carbonate (H-Carb) of each ionic liquid composed of [$\mathrm{N2222^+}$] or [$\mathrm{N4444^+}$] cations combined independently with each of the three halide anions (6 different ILs) is studied. Experimentally, ILs with [$\mathrm{N2222^+}$] cation were insoluble in the reaction medium, while [$\mathrm{N4444^+}$]-based ILs are soluble \cite{hernandez2024dissociation}. Experimental catalytic activity relation with the cation and halide anion selection was investigated for [$\mathrm{N4444^+}$], [$\mathrm{bmim^+}$], and [$\mathrm{emim^+}$] cations combined with the three halide anions and for [$\mathrm{N1111^+}$][$\mathrm{Br^-}$] (10 ILs confirmed experimentally to be soluble). Experimental catalytic activity for a common cation and the three halides followed the order [$\mathrm{I^-}$] $>$ [$\mathrm{Br^-}$] $>$ [$\mathrm{Cl^-}$]. In terms of the cation influence for fixed [$\mathrm{Br^-}$] halide, experimental yields displayed the order [$\mathrm{N4444^+}$] $>$ [$\mathrm{bmim^+}$] $>$ [$\mathrm{emim^+}$] $>$ [$\mathrm{N1111^+}$] \cite{hernandez2024dissociation}. Therefore, the general objective of this work is to find the underlying mechanisms by which ions influence the catalytic activity of halide-based ionic liquid catalysts in the formation of hexylene carbonate (H-Carb) from CO$_{2}$ cycloaddition to 1,2-epoxyhexane, using molecular dynamics as the simulation method.

\section{Simulation details}
\label{sec:sample2}

The MD simulations performed in this work can be divided into two main groups. The first group is dedicated to epoxides and carbonates force fields testing and validation by the determination of the vapor-liquid coexistence densities. The second group is dedicated to the ILs interaction with 1,2-epoxyhexane and hexylene carbonate. All simulations have been performed with GROMACS (version 2016.5, double precision) software. Spatial Distribution Functions (SDFs) were computed with TRAVIS software \cite{brehm2011travis, brehm2020travis}. \\

\subsection{Vapor-liquid equilibria of pure compounds}

Several simulation methods for vapor-liquid equilibrium properties determination have been proposed in the literature. Popular methods involving chemical potentials determination like Widom's test particle insertion, are well-suited for systems of small molecules but become impractical for densely packed liquid phases where molecular insertion is hindered. On the other hand, recent works performed in the grand canonical ensemble have been demonstrated to be accurate for systems including big particles or even polar molecules \cite{khani2022water, gharibi2024self}. An additional method corresponds to the direct coexistence simulation technique, which permits the simultaneous simulation of the desired phases together \cite{blas2008vapor}. Given that the molecules employed in this section correspond to complex organic epoxides and carbonates with polar regions coupled with alkyl side chains, the direct coexistence technique was selected among the alternatives to perform a qualitative evaluation of the force fields employed in the subsequent sections of this work. Vapor-liquid equilibrium is therefore studied through MD simulations at different temperatures in the canonical ensemble ($NVT$) using
the direct coexistence simulation technique. Following this method, the vapor and liquid phases are put together in the same simulation box, which allows to determine both vapor and liquid coexistence densities once the system has reached the equilibrium. The initial configuration of the liquid phase was generated by random molecule insertions into the simulation box using PACKMOL software package. The box dimensions and the number of molecules ($N$) used in this work are presented in Table \ref{table_1}. The length of the simulation box along the $x$- and $y$--axis ($L_{x}$ and $L_{y}$) was set greater than $10\sigma$ to ensure that the minimum image convention (MIC) was satisfied for periodic boundary conditions (PBC) utilization. Here $\sigma$ is the largest diameter parameter value of the LJ intermolecular potential of the compounds considered in this work. The $z$-axis length ($L_{z}$) was selected to obtain a wide enough liquid slab to ensure that the two vapor-liquid interfaces generated afterward could be considered independent from each other. Initial bulk liquid configurations were minimized and stabilized first by running a short $NVT$ simulation (2 ns) using the velocity-rescaling (v-rescale) thermostat and then by a short $NP_zT$ simulation (2 ns) in which pressure was controlled by a semi-isotropic Parrinello-Rahman (20 bar) barostat, allowing only fluctuations of the simulation box along the $z$--axis with a compressibility factor value of $4.5 \times 10^{-5}\,\text{bar}^{-1}$. Finally, an expansion in $L_{z}$ was performed, creating a space that can be occupied by some molecules of the liquid phase forming the coexistence vapor phase. The final dimensions of the initial simulation boxes used for determining the vapor-liquid phase equilibria of the propylene oxide (P-Epox), propylene carbonate (P-Carb), and 1,2-epoxyhexane (H-Epox) are summarized in Table \ref{table_1}. Finally, vapor-liquid $NVT$ simulations for each selected temperature were carried out using the v-rescale thermostat and the Verlet leapfrog algorithm with a timestep of 1 fs. Simulations were run for  10 ns. In all cases, a $r_c$ value of 1.79 nm ($r_{c}=5\sigma$) was used without long-range corrections for the Lennard-Jones (LJ) dispersive interactions and using the particle mesh Ewald (PME)\cite{Essmann1995a} long-range correction for the Coulombic interactions.  

The LJ and Coulombic model parameters of the P-Epox and P-Carb force fields were found on the Automatic Topology Builder (ATB) public website \cite{malde2011automated, stroet2018automated}. 1,2-epoxyhexane (H-Epox) and hexylene carbonate (H-Carb) (compounds present in the reference experimental results \cite{hernandez2024dissociation}) force fields were constructed by the combination of two force fields. For H-Epox, P-Epox, and butyl cyclopropane force fields (from ATB \cite{malde2011automated, stroet2018automated}) were combined. H-Carb was correspondingly constructed by the combination of P-Carb and butyl cyclopentane force fields (from ATB \cite{malde2011automated, stroet2018automated}). Atomic charges were re-parametrized after the combination to ensure zero net charge using NBO over the corresponding DFT-optimized geometry obtained by using BP86 functional, coupled with a TZVP basis set, which corresponds to the same parameters that have been successfully used for other applications of these compounds \cite{hernandez2024dissociation, hernandez2022universal}. In all cases, the non-bonded interaction parameters between unlike groups were calculated through the geometric-mean Lorentz-Berthelot combination rules.

\begin{table}
    \centering
    \caption{Initial simulation box dimensions and number of molecules used to study the vapor-liquid equilibria of the propylene oxide (P-Epox), propylene carbonate (P-Carb), and 1,2-epoxyhexane (H-Epox) through $NVT$ MD simulations.}
    \begin{tabular}{lccccccccc}
    \hline
    \hline
         & \,\, & $L_{x=y}^{\text{box}}$ (nm) & \,\, &  $N$ & \,\, & $L_{z}^{\text{liquid}}$(nm) & \,\, & $L_{z}^{\text{box}}$(nm) \\
    \hline
        P-Epox & \,\, & 3.6 & \,\, & 1325 & \,\, & 10.8 & \,\, & 30 \\
        P-Carb & \,\, & 3.6 & \,\, & 825 & \,\, & 10.8 & \,\, & 36 \\
        H-Epox & \,\, & 4.0 & \,\, & 1040 & \,\, & 16.8 & \,\, & 40 \\

    \hline
    \hline        
    \end{tabular}
    \label{table_1}
\end{table}

\subsection{Binary mixture simulations involving ionic liquids}

Currently, several methods have been demonstrated in the literature to be accurate and efficient in predicting phase behaviour by chemical potentials calculations. Successful techniques applied to complex systems include Hamiltonian scaling \cite{errington1998phase}, test-particle-insertion in polymers \cite{van1996free} or adaptive-numerical-bias
metadynamics in three-dimensional nanocrystals \cite{eslami2024metadynamics}. On the other hand, the direct coexistence technique has been successfully applied to complex systems like CO$_{2}$ encapsulation in hydrates \cite{miguez2015molecular} and, similarly to this work, in CO$_{2}$ conversion to cyclic carbonates in porous ionic liquids \cite{clark2023solvation}. Therefore, ionic liquids simulations performed in this work are divided into two sub-groups, semi-isotropic
and isotropic $NPT$ simulations. Semi-isotropic $NPT$ simulations were constructed for the solubility analysis of ionic liquids in both 1,2-epoxyhexane (H-Epox) and hexylene carbonate (H-Carb) together. Each simulation contains an ionic liquid composed of an anion and a cation combination in addition to the H-Epox and H-Carb. A total of six simulations were carried out, which correspond to two cations ([$\mathrm{N4444^+}$] and [$\mathrm{N2222^+}$]) each one combined independently with the three halide anions ([$\mathrm{I^-}$], [$\mathrm{Br^-}$], and [$\mathrm{Cl^-}$]). The initial configuration was constructed by the combination of three pre-simulated and stabilized systems, one for the ionic liquid, another for the epoxide, and another for the carbonate. The pre-simulation process consisted of several steps, applied to each of the pure phases. The process for each pure phase started with a minimization of the randomly generated configuration. The next step was a semi-isotropic $NPT$ simulation in which a packed liquid configuration was obtained. Then, the three pure configurations were unified in a single simulation box, obtaining a configuration that was again minimized before the simulation. Geometric-mean Lorentz-Berthelot combination rule was employed in all cases. Simulations were run using the velocity Langevin dynamics integrator (sd), which adds a friction and a noise term to Newton’s equations of motion to control the system temperature. In order to sample correctly the flexibility of the molecules, a time step of 1 fs was employed.  Simulations were run until the system reached the equilibrium ($\approx$500 ns depending on the cation). The $r_c$ value was selected to be equal to 1.35 nm for both LJ and Coulomb interactions. For the LJ interactions, long-range corrections were applied. Also, Coulomb interaction corrections were performed by the fast smooth particle-mesh Ewald method (PME).~\cite{Essmann1995a} Notice that the use of the PME corrections is strictly necessary since they are set by default in the ionic liquid CL\&P model, which is the force field model utilized for all the ionic liquids in this work \cite{bernardes2017modeling, canongia2012cl}. The experimental reaction conditions of temperature and pressure were set to 65ºC and 20 bar by the sd integrator (with a friction constant of 1 ps$^{-1}$) and the Parrinello-Rahman barostat (with a time constant of 0.5 ps and a compressibility factor along the $z$-axis of $4.5\times10^{-5}\,\text{bar}^{-1}$) respectively. The number of molecules of each component corresponds to 500 H-Epox and 250 H-Carb for the organic species. For [$\mathrm{N4444^+}$] cation-based ILs, 200 molecules were introduced, while for [$\mathrm{N2222^+}$] cation-based ILs, due to the reduced number of atoms for the shorter alkyl side chains of the cation, 250 molecules could be employed. \\

Isotropic $NPT$ simulations were performed aiming to analyze the obtained radial and spatial distribution functions (RDFs and SDFs) between several atom pairs that were considered crucial for the reaction process in the literature. For these simulations, ionic liquids and H-Epox were the only species used, excluding H-Carb from the analysis to center the analysis on the catalyst-reagent interaction. Simulation boxes were constructed by random molecule insertions of 200 ILs (200 cations and 200 anions) in an oversized simulation box. Then 1,2-epoxyhexane (H-Epox) molecules were randomly inserted (600 H-Epox molecules) in the configuration containing the ionic liquid. Lastly, the resulting configuration was minimized. This procedure was followed to obtain a mixed configuration in which diffusion would have already occurred, skipping the mixing step that was found to be slow in the solubility analysis. The simulation details for the barostate and thermostat utilized in these simulations are those specified previously for the ionic liquids semi-isotropic $NPT$ simulations, but the barostate was applied isotropically. A time step of 0.001 ns was employed, simulations were run for 60 ns, and an expanded $r_c$ value of 2.7 nm was used. A total of ten simulations were performed, corresponding to each ionic liquid resulting from the combination of [$\mathrm{N4444^+}$], [$\mathrm{bmim^+}$], and [$\mathrm{emim^+}$] cations with the three halide anions and [$\mathrm{N1111^+}$][$\mathrm{Br^-}$] IL. The 10 corresponding ILs were found to be soluble experimentally \cite{hernandez2024dissociation}, which is the reason  for the [$\mathrm{N2222^+}$] cation exclusion in these simulations and the [$\mathrm{N1111^+}$][$\mathrm{Br^-}$] inclusion.

\section{Results and Discussion}
\label{sec:results}

\subsection{Vapor-Liquid phase equilibrium of Epoxides and Carbonate}
\label{sec:liquid-vapor}

Propylene oxide (P-Epox) and carbonate (P-Carb) force fields have been evaluated in terms of their pure vapor-liquid equilibrium properties by the direct coexistence method. The visual inspection of the P-Epox $NVT$ simulation boxes with temperatures 200 and 400 K can be observed in Fig.\ref{fig:Visual_P-Epox} (See \ref{sec:sample:appendix}). Similar representations were obtained for the rest of the systems. Fig.\ref{fig:Visual_P-Epox} shows that at 200 K there is no vapor phase in coexistence with the liquid for P-Epox. Therefore, the liquid phase remains packed in the center of the box with negligible vapor pressure. On the other hand, at 400 K a vapor phase is visible in the form of moving molecules through the whole $z$ axis. Additionally, liquid phases occupy a greater extension. This means that the vapor phase is composed of fewer molecules that occupy a greater volume, which is why a reduced density is expected to be found. \\

\begin{figure}[h!]
    \centering
    \includegraphics[width=0.7\linewidth]{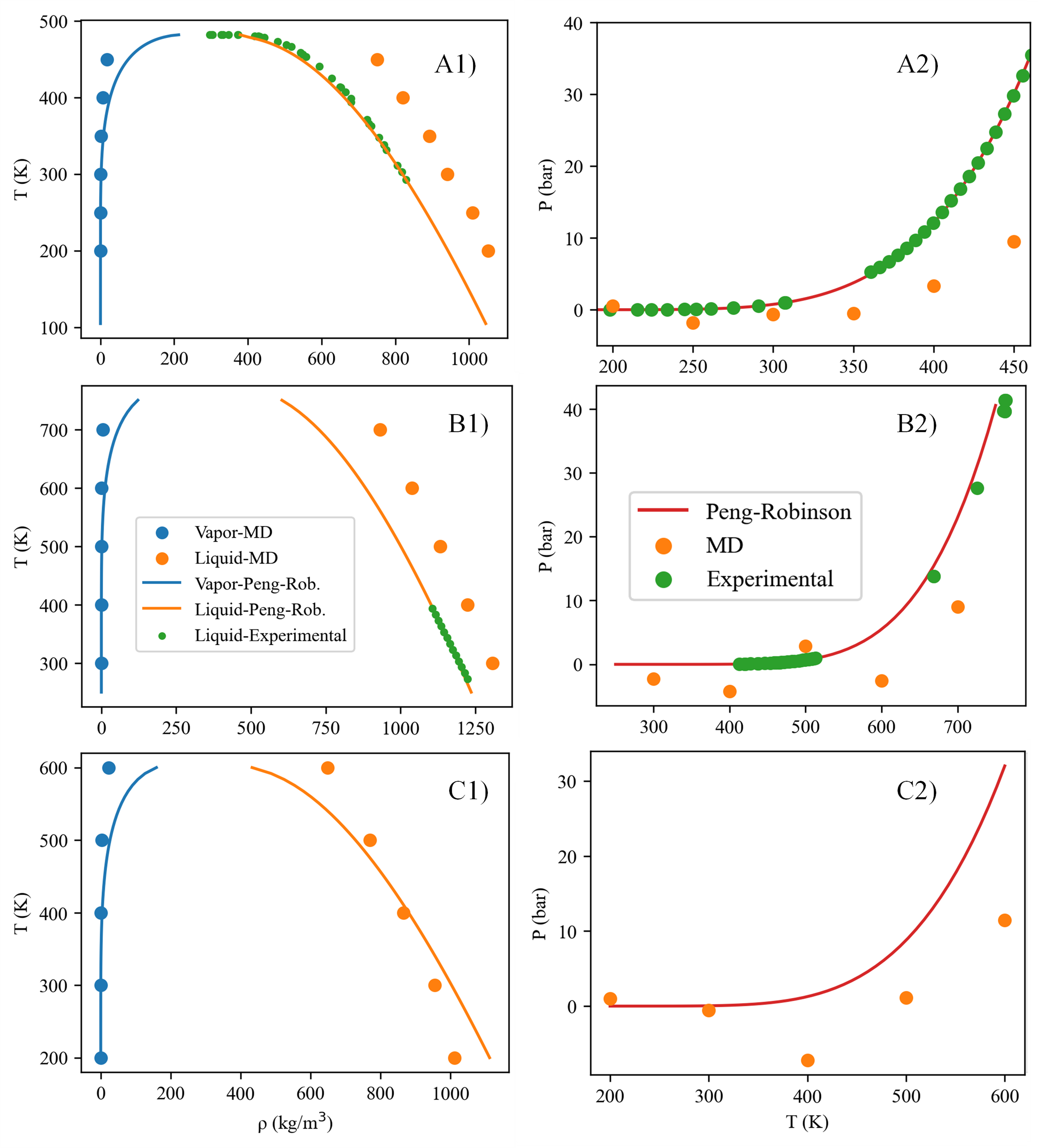}
    \caption{$T-\rho$ and $P-T$ phase equilibrium diagrams obtained by MD and PR. EoS. A) P-Epox compared with experimental data \cite{rutenberg1973pvt, kobe1956critical,frenkel2006nist}  B) P-Carb compared with experimental data \cite{stull2002vapor, gurevich1982viscosity, chernyak2004vapor, wilson2002critical} and C) H-Epox. Experimental data and PR predictions were obtained through the ASPEN-NIST properties engine.}
    \label{fig:coex-todos}
\end{figure}

Equilibrium vapor-liquid properties for P-Epox and P-Carb will first be analyzed. The density profiles obtained can be found in Fig.\ref{fig:1SUB}A and Fig.\ref{fig:1SUB}B respectively in the \ref{sec:sample:appendix}. In these figures, it can be seen that for greater temperatures, liquid densities (center top values of the curves) find a lower value, while gas densities (right and left curve bases) increase by displacing in the vertical axis. Additionally, interphase thicknesses get bigger as the temperature is increased. From each curve in Fig.\ref{fig:1SUB}, a liquid and a vapor density for each temperature and compound can be obtained through the average of the values encountered at each vapor and liquid phase, enabling the generation of the $T-\rho$ diagrams. $P-T$ phase diagrams can also be generated by attending to the transversal pressure of the simulation boxes along the $z$-axis, which is the direction in which the vapor-liquid interface is generated. The V-L equilibrium properties obtained by molecular dynamics (MD) simulations, along with the experimental values found for the liquid phases, were gathered and compared with the Peng-Robinson Equation of State (EoS) predictions for the liquid, being this model used afterward for the gas density prediction. Results can be seen in Fig.\ref{fig:coex-todos}A for propylene oxide (P-Epox), in Fig.\ref{fig:coex-todos}B for propylene carbonate (P-Carb), and in Fig.\ref{fig:coex-todos}C for 1,2-epoxyhexane (H-Epox). Attending to Fig.\ref{fig:coex-todos}A1 and Fig.\ref{fig:coex-todos}B1, it can be noted that experimental data is in great agreement with Peng-Robinson predictions for the liquid phase. When comparing MD densities with those predicted by Peng-Robinson, vapor densities are slightly underestimated, especially at 400 and 450 K for P-Epox and at 700 K for P-Carb. Correspondingly, liquid densities are overestimated for both compounds by $150\,\text{kg/m}^3$. The force field models are then capable of reproducing the density trend visible in the phase diagram, even though the density values are displaced from the expected ones. Additionally, the critical points appear to be overestimated by the shape of the MD density curves for both P-Epox and P-Carb. In terms of the $P-T$ phase diagrams visible in Fig.\ref{fig:coex-todos}A2 and Fig.\ref{fig:coex-todos}B2, it was found that Peng-Robinson perfectly describes the $P-T$ curves, while MD poorly follows an approximate trend for P-Epox with a slight improvement for P-Carb, being pressures underestimated in both cases due to the few molecules existing in the gas phase. Nevertheless, and because the results are qualitatively correct, both force fields were used in the construction of corresponding H-Epox and H-Carb force fields. The 1,2-epoxyhexane (H-Epox) force field was constructed as the combination of the P-Epox and butyl cyclopropane force fields along with atom charges reparametrization to the values obtained after DFT geometry optimization to ensure net zero charges. The resulting force field is evaluated following the same methodology employed previously for P-Epox and P-Carb. Density profiles were used to obtain the $V-L$ coexistence densities. The obtained $T-\rho$ and $P-T$ phase diagrams can be seen in Fig.\ref{fig:coex-todos}C1 and Fig.\ref{fig:coex-todos}C2 respectively, compared only with Peng-Robinson predictions since no experimental data could be found. The results (see Fig.\ref{fig:coex-todos}C) show the great impact of the changes performed to the P-Epox to obtain the H-Epox force field. If Peng-Robinson predictions are correct (which they were accurate for P-Epox and P-Carb), gas densities remain underestimated at the highest temperatures (500 and 600 K), while liquid densities are underestimated at 200 and 300 K and overestimated at 500 and 600 K, being the density value at 400 K almost equal to that of Peng-Robinson prediction. Additionally, the critical point is overestimated given that Peng-Robinson prediction situates it in the vicinity of 600 K. These results contrast with those obtained for the P-Epox and P-Carb force fields, in which liquid density values were displaced by $150\,\text{kg/m}^3$. The H-Epox densities seem to be numerically closer to Peng-Robinson predictions. The interval comprehended between 300 and 500 K is the one that better reproduces densities, which corresponds to the experimental conditions (65 ºC) that will be reproduced in the following sections containing the ionic liquid, H-Epox, and H-Carb molecules. Figure \ref{fig:coex-todos}C1 shows, like in the previous homologous diagrams, that the $P-T$ behavior is poorly reproduced with a lot of uncertainty for reduced pressures, probably as a consequence of the small number of molecules reaching the gas phase. Nevertheless and because the best-expected results are contained in the objective temperature interval, the constructed force field models are used in the subsequent sections of this work.

\subsection{Ionic Liquids Solubility Analysis in 1,2-epoxyhexane (H-Epox) and hexylene carbonate (H-Carb)}
\label{sec:Solubility Analysis}

Ionic liquid solubilities in 1,2-epoxyhexane (H-Epox) and hexylene carbonate (H-Carb) have been studied for ionic liquids composed of [$\mathrm{N2222^+}$] and [$\mathrm{N4444^+}$] cations, independently combined with the three halide anions [$\mathrm{I^-}$], [$\mathrm{Br^-}$], and [$\mathrm{Cl^-}$]. This constitutes a total of six semi-isotropic $NPT$ simulations at the experimental conditions (65ºC and 20 bar) with three initial independent phases of H-Epox, IL, and H-Carb distributed along the $z$-axis. The visual inspection for the final configuration of the simulation corresponding to the [$\mathrm{N4444^+}$][$\mathrm{Cl^-}$] is shown in Fig.\ref{fig:todos-solu} (see \ref{sec:sample:appendix}). In this figure, it can be seen that the H-Epox and H-Carb are mutually miscible. For a better visualization of the ILs molecules to analyze ILs solubilities, H-Epox and H-Carb have been excluded from the following visual inspections presented in this section.  \\

\begin{figure}[h!]
    \centering
    \includegraphics[width=0.8\linewidth]{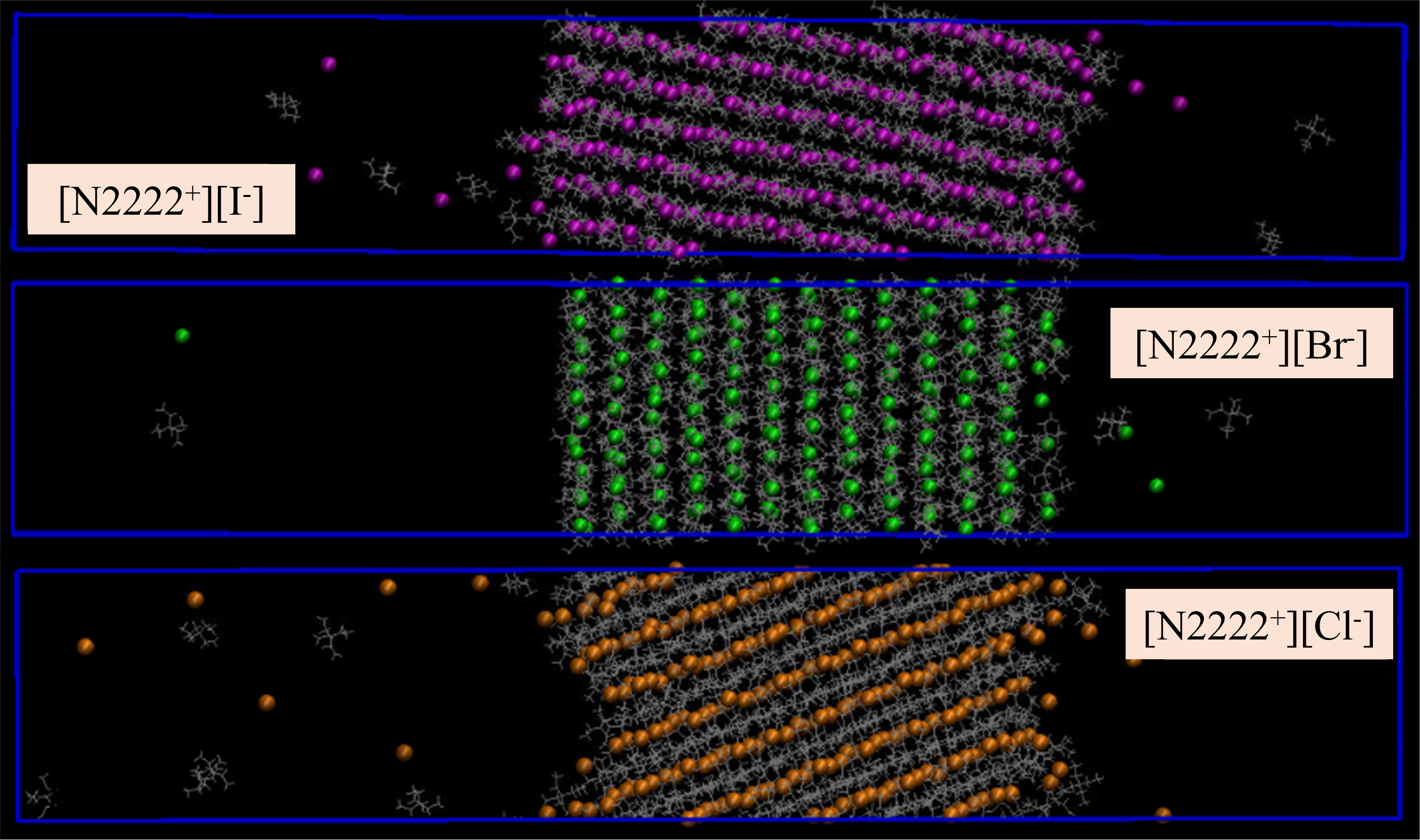}
    \caption{Visual VMD representation of the semi-isotropic $NPT$ simulations of [$\mathrm{N2222^+}$] [I/Br/Cl], H-Epox, and H-Carb. The last two compounds are excluded from the representation for better visualization of the ionic liquid.}
    \label{fig:N2222-solu}
\end{figure}

The visual inspection of the final configurations of ILs composed of [$\mathrm{N2222^+}$] cation combined with [$\mathrm{I^-}$], [$\mathrm{Br^-}$], and [$\mathrm{Cl^-}$] anions, excluding H-Epox and H-Carb from the representation can be seen in Fig. \ref{fig:N2222-solu}. This figure shows that the three ionic liquid systems have crystallized. Cations and anions are therefore distributed in regular structures, which implies a low or negligible solubility of these ILs into the epoxide or the carbonate. This result constitutes a validation of the molecular models employed in this work, as the three ionic liquids were experimentally determined to be insoluble \cite{hernandez2024dissociation} in H-Epox and H-Carb. Experimental results found very low carbonate yields for these ILs, varying from 1\% for [$\mathrm{N2222^+}$][$\mathrm{I^-}$] to 4\% for [$\mathrm{N2222^+}$][$\mathrm{Cl^-}$] \cite{hernandez2024dissociation}. Therefore, it becomes clear that, for insoluble ILs, epoxide conversions will be low. The catalyst is not able to properly interact with the epoxide for reaction, a situation that was supported experimentally and, now, theoretically confirmed by the results obtained in this work \cite{hernandez2024dissociation}.

\begin{figure}[h!]
    \centering
    \includegraphics[width=0.8\linewidth]{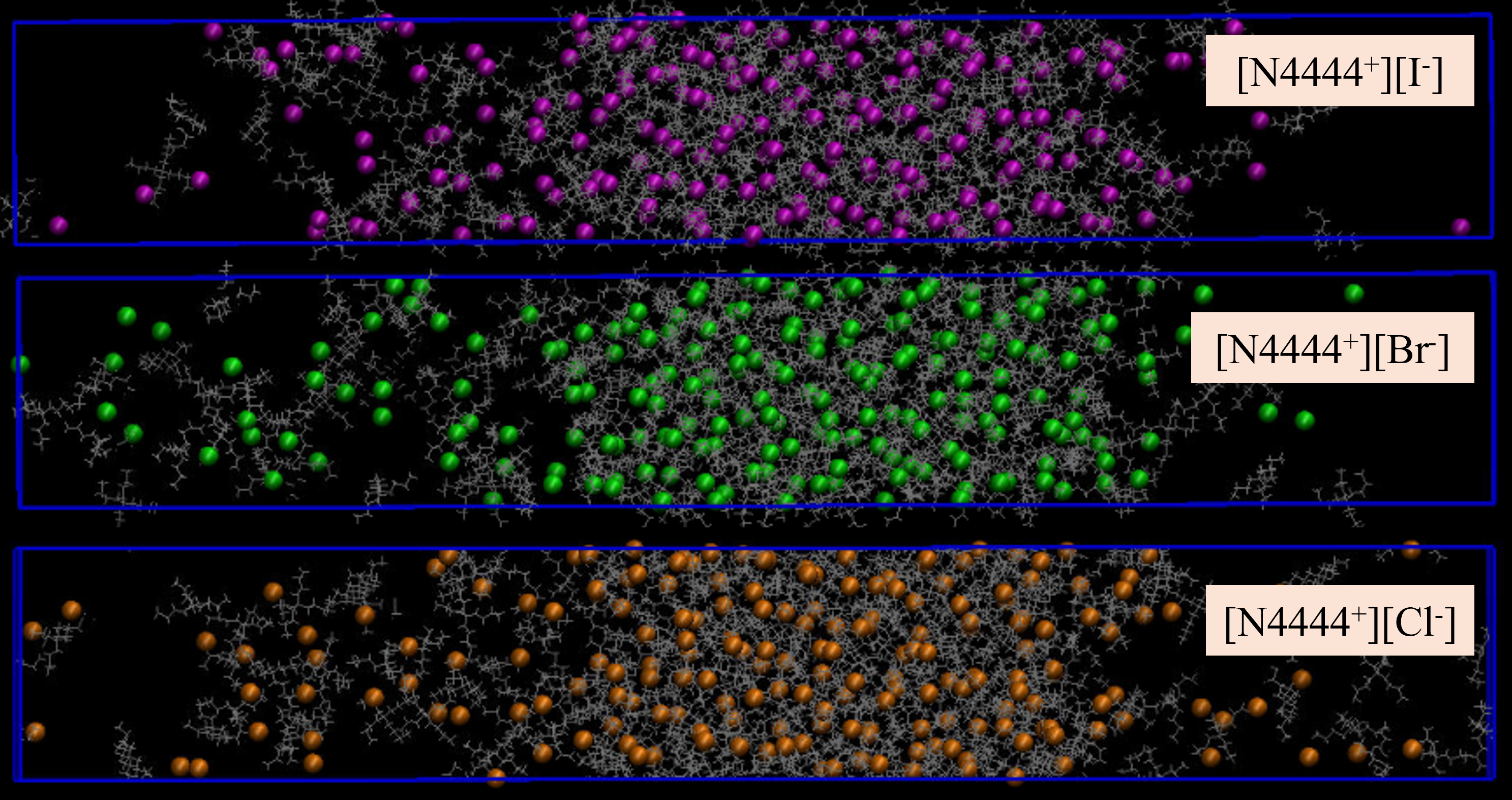}
    \caption{Visual VMD representation of the semi-isotropic $NPT$ simulations of [$\mathrm{N4444^+}$] [I/Br/Cl], H-Epox and H-Carb. The last two compounds are excluded from the representation for better visualization of the ionic liquid.}
    \label{fig:N4444-solu}
\end{figure}

The visual inspection of the final configurations for ILs composed of the [$\mathrm{N4444^+}$] cation combined with the three different halides can be observed in Fig.\ref{fig:N4444-solu}. H-Epox and H-Carb were again excluded from the representation to achieve a better visualization. Attending to Fig.\ref{fig:N4444-solu}, ILs phases remain disordered after the simulations and with clear diffusion of the ionic liquids in the H-Epox and H-Carb medium. Therefore, crystallization has not happened, meaning that these ionic liquids are miscible in the H-Epox and H-Carb phases. This constitutes another validation of the employed force fields, as experimental results showed that [$\mathrm{N4444^+}$][$\mathrm{Br^-}$], [$\mathrm{N4444^+}$][$\mathrm{Cl^-}$], and [$\mathrm{N4444^+}$][$\mathrm{I^-}$] are all soluble at the end of the reaction \cite{hernandez2024dissociation}, which is the case in these simulations with the H-Carb presence. System dynamics were slow, which summed to the excessive concentration of ionic liquid (used to determine solubility) is the reason why a homogeneous distribution of the ionic liquids (clearly visible for [$\mathrm{N4444^+}$][$\mathrm{Cl^-}$]) needed of long simulation times. Therefore, in terms of solubility, simulations performed with [$\mathrm{N2222^+}$] cation eventually crystalized for the three halide anions, while those performed with [$\mathrm{N4444^+}$] cation with the same halide anions did not. The results therefore align with the literature, indicating that the cation predominantly determines the solubility of the ionic liquids \cite{bobbink2018intricacies, kawanami2003rapid}. Specifically, it is the side chain lengths the element that most affects the change of the solubility. Longer alkyl side chains improve the ILs solubility, at least in an H-Epox and H-Carb medium, which has an enormous impact on the catalytic activity since soluble ionic liquids are expected to be far more reactive with epoxide than insoluble ones. This is in fact what was found experimentally \cite{hernandez2024dissociation} with carbonate yields varying from 91\% from [$\mathrm{N4444^+}$] [$\mathrm{I^-}$] to 1\% for [$\mathrm{N2222^+}$] [$\mathrm{I^-}$] and confirmed now by MD simulations for both [$\mathrm{N2222^+}$] and [$\mathrm{N4444^+}$] cations combined with the three halide anions. \\

Finally, by observing Fig.\ref{fig:N2222-solu} for the [$\mathrm{N2222^+}$] case, it appears that even though anions mobility should be superior to cations as they are single-atom ions, the diffused number of cations and anions is somehow equivalent in these simulations. In fact, this may have greater implications than expected, as solubility might be dominated by the more voluminous cation's diffusion capability, while halide anions may follow diffused cations forming either ion pairs or aggregated species in some proportion, being the first case visible in Fig.\ref{fig:N2222-solu} representations. By expanding this analysis for the [$\mathrm{N4444^+}$] simulations (Fig.\ref{fig:N4444-solu}), the diffused cation-anion distribution resembles the one introduced for [$\mathrm{N2222^+}$] cation-based ILs. If halide anions were expected to have greater mobility than [$\mathrm{N2222^+}$] cation, it should be even more notorious for [$\mathrm{N4444^+}$] cation in which side alkyl chains have lengthened in two carbons each. But again, attending to Fig.\ref{fig:N4444-solu}, a homogeneous distribution of the halide anions is not found. It looks like it is cations diffusion that widens the halide anion's mobility range, forming ion-pairs and aggregates instead of dissociated species. This is in great correspondence with the low conductivities found experimentally, or the also low dissociation constants of ILs in epoxides and carbonates computed with DFT/COSMO-RS \cite{hernandez2024dissociation}. The results found in these simulations point in the direction that independent cations and anions should not be considered the main reactive species in these systems, as their presence is very reduced and does not easily explain conversions of up to 90\%. The ionic pairs may be the true catalyzers of this reaction, or the presence of a coordinating species (as already proposed in the literature \cite{clark2023solvation}) must be required.

\subsection{Reaction Medium Interactions}
\label{sec:Reaction Medium Interactions}

\begin{figure}[h]
    \centering
    \includegraphics[width=0.8\linewidth]{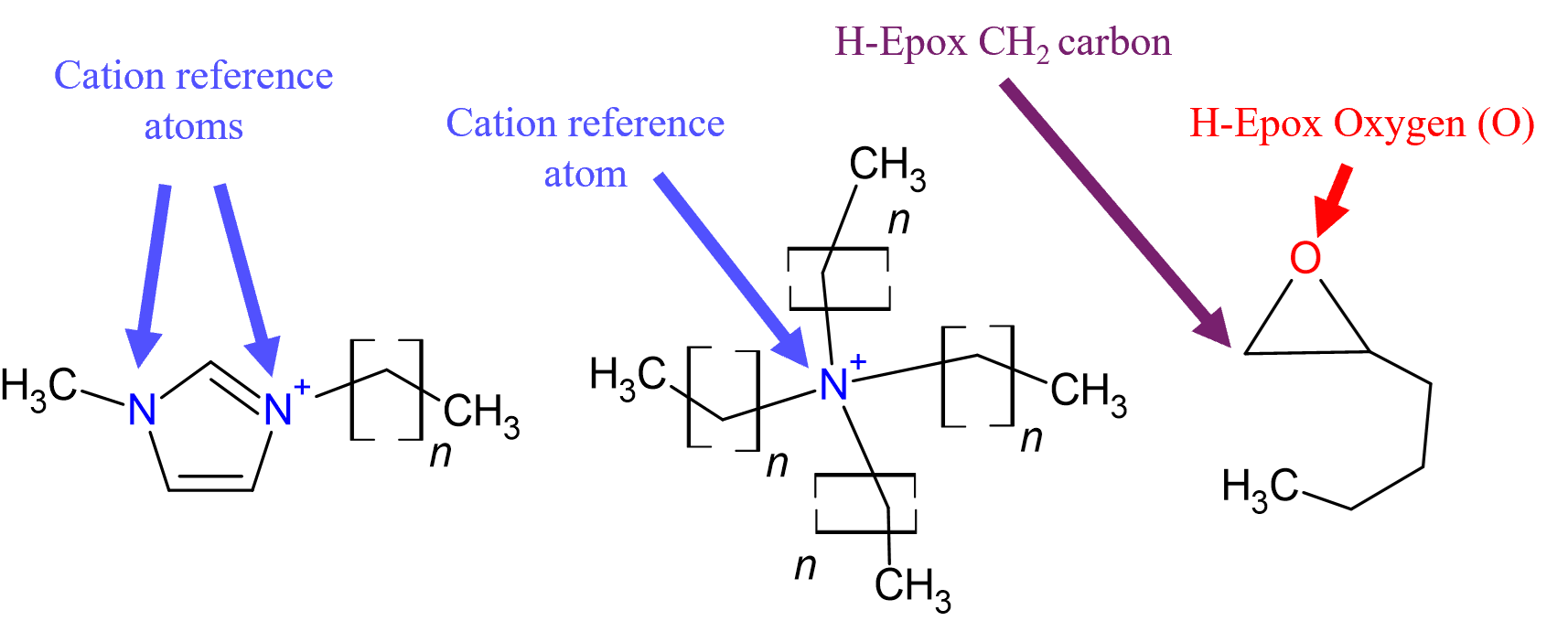}
    \caption{Structures, reference atoms, and nomenclature employed for Radial Distribution Functions (RDFs) computation.}
    \label{fig:REFATMS}
\end{figure}

In this section, a better understanding of the catalytic activity displayed by halide anion variation and the cation effect will be searched. A total of 10 systems are studied in this section, each one composed of the mixture of 1,2-epoxyhexane (H-Epox) and each of the possible ILs resulting from the combination of each the cations; [$\mathrm{N4444^+}$], [$\mathrm{bmim^+}$], and [$\mathrm{emim^+}$] with the three halide anions ([$\mathrm{Cl^-}$], [$\mathrm{Br^-}$] and [$\mathrm{I^-}$]), and [$\mathrm{N1111^+}$][$\mathrm{Br^-}$] IL (all confirmed experimentally to be soluble \cite{hernandez2024dissociation}). The analysis was performed in terms of the reference atoms shown in Fig.\ref{fig:REFATMS}, which are thought to be decisive in the reaction mechanism. \\

\begin{figure}[h!]
    \centering
    \includegraphics[width=1\linewidth]{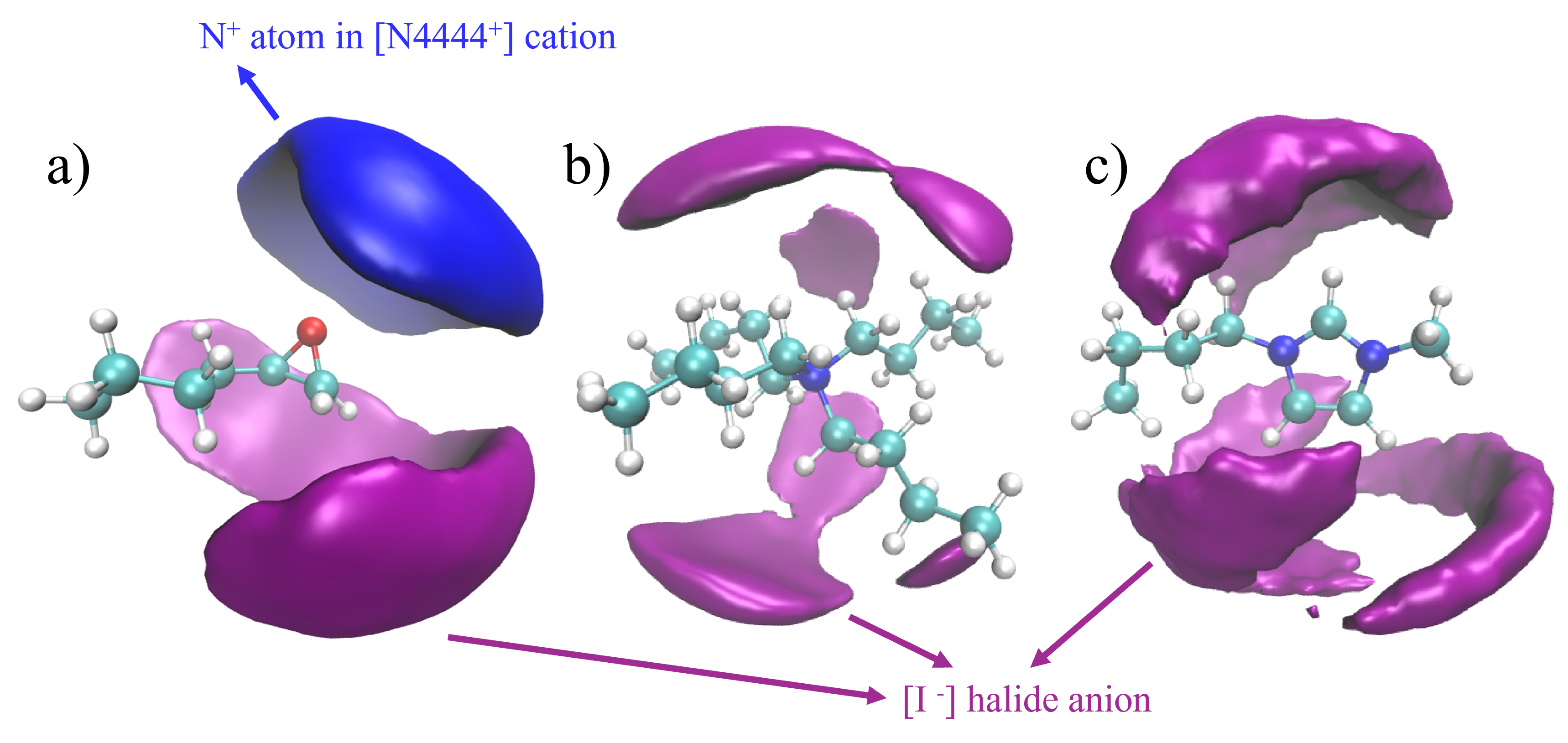}
    \caption{Spatial Distribution Functions (SDFs). a) Nitrogen atom of the cation and [$\mathrm{I^-}$] halide anion with respect to the 1,2-epoxyhexane of the [$\mathrm{N4444^+}$][$\mathrm{I^-}$]+1,2-epoxyhexane simulation. b) [$\mathrm{I^-}$] halide anion with respect to the [$\mathrm{N4444^+}$] cation of the [$\mathrm{N4444^+}$][$\mathrm{I^-}$]+1,2-epoxyhexane simulation. c) [$\mathrm{I^-}$] halide anion with respect to the [$\mathrm{bmim^+}$] cation of the [$\mathrm{bmim^+}$][$\mathrm{I^-}$]+1,2-epoxyhexane simulation.}
    \label{fig:SDFS}
\end{figure}

Specific Spatial Distribution Functions (SDFs) for [$\mathrm{N4444^+}$][$\mathrm{I^-}$] and [$\mathrm{bmim^+}$][$\mathrm{I^-}$] simulations can be seen in Fig.\ref{fig:SDFS}. This figure is composed of a set of three SDFs. First two: a) and b) were obtained from [$\mathrm{N4444^+}$][$\mathrm{I^-}$]+1,2-epoxyhexane simulation, while c) corresponds to the homologous [$\mathrm{bmim^+}$][$\mathrm{I^-}$]+1,2-epoxyhexane simulation. By first analyzing Fig. \ref{fig:SDFS}a), the distribution of both the [$\mathrm{N4444^+}$] cation and the [$\mathrm{I^-}$] halide anion can be understood relative to the H-epox molecule. It can be seen that [$\mathrm{N4444^+}$] cation can be more frequently found surrounding the oxygen atom of the epoxide, in the opposite direction of the epoxide chain. Additionally, it can be noted that [$\mathrm{I^-}$] situates itself wrapping both carbon atoms of the epoxide ring. This result reveals how the cation could be activating the epoxide for the subsequent halide nucleophilic attack for ring opening, or a coordinated effect of both ions could be happening. It can be seen that the epoxide chain prevents the ions presence. This effect would vary for different epoxide substituents, generating possible steric hindrances that could reduce or widen the space at which the reaction is permitted. By analysing Fig.\ref{fig:SDFS}b) and c), the difference between [$\mathrm{I^-}$] anion distribution around [$\mathrm{N4444^+}$] and [$\mathrm{bmim^+}$] cations can be understood. [$\mathrm{I^-}$] anion distributes around [$\mathrm{N4444^+}$] in a spherical-like shape around the closest hydrogen atoms to the nitrogen central atom. On the other hand , for [$\mathrm{bmim^+}$] cation, [$\mathrm{I^-}$] anion distributes around the hydrogen atoms of the [$\mathrm{bmim^+}$] closest to both nitrogen atoms of the cation. \\

Radial Distribution Functions ($g(r)$) are computed in terms of the reference atoms or atom groups shown in Fig. \ref{fig:REFATMS}. The three types of computed RDFs for each ionic liquid system correspond to I) Cation and Anion (C-A) which is calculated between the cations reference atoms (see Fig. \ref{fig:REFATMS}) and the respective halide anion. II) Epoxide (CH$_2$) carbon and Anion (Epox-A) calculated between the CH$_2$ carbon in the H-Epox and the halide anion. III) Cation and epoxide (C-Epox) calculated between the cation reference atoms and the oxygen in H-Epox. A total of 10 simulations were performed, from which 30 RDFs were obtained.

\begin{figure}[h!]
    \centering
    \includegraphics[width=0.8\linewidth]{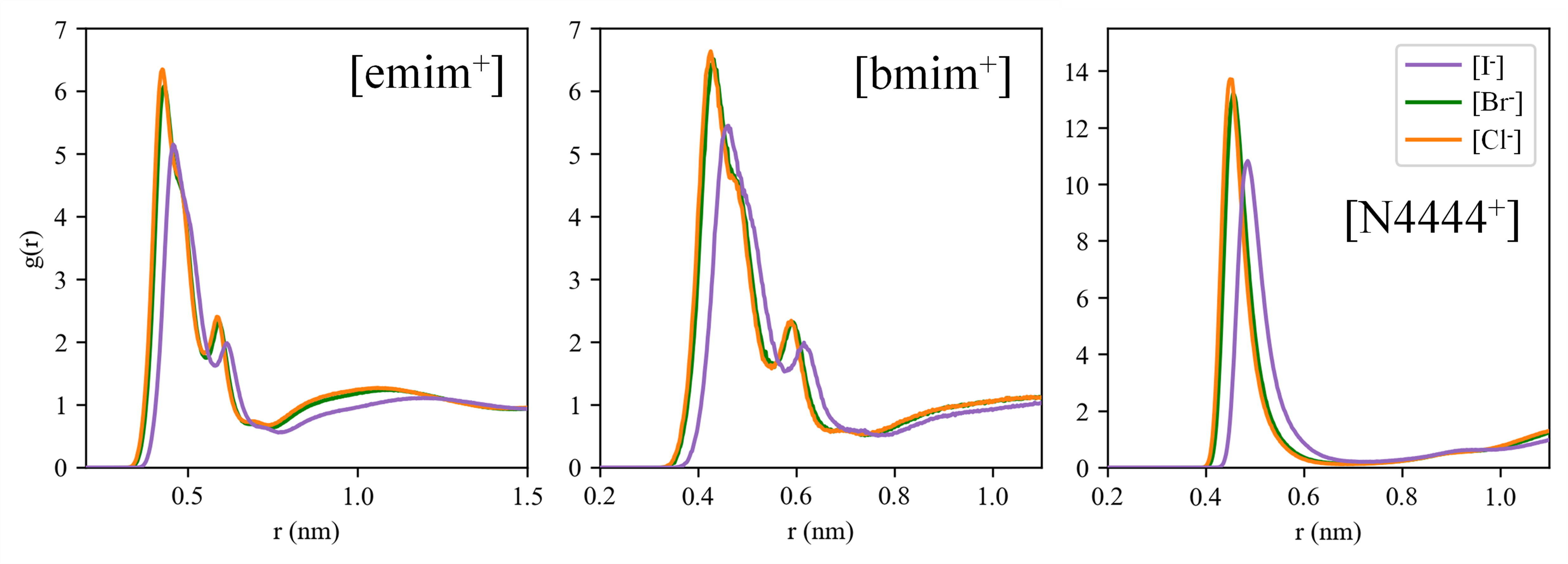}
    \caption{Cation-anion (C-A) radial distribution functions (RDFs) obtained for each cation and halide anion combination.}
    \label{fig:RDFS-CA}
\end{figure}

\begin{figure}[h!]
    \centering
    \includegraphics[width=0.8\linewidth]{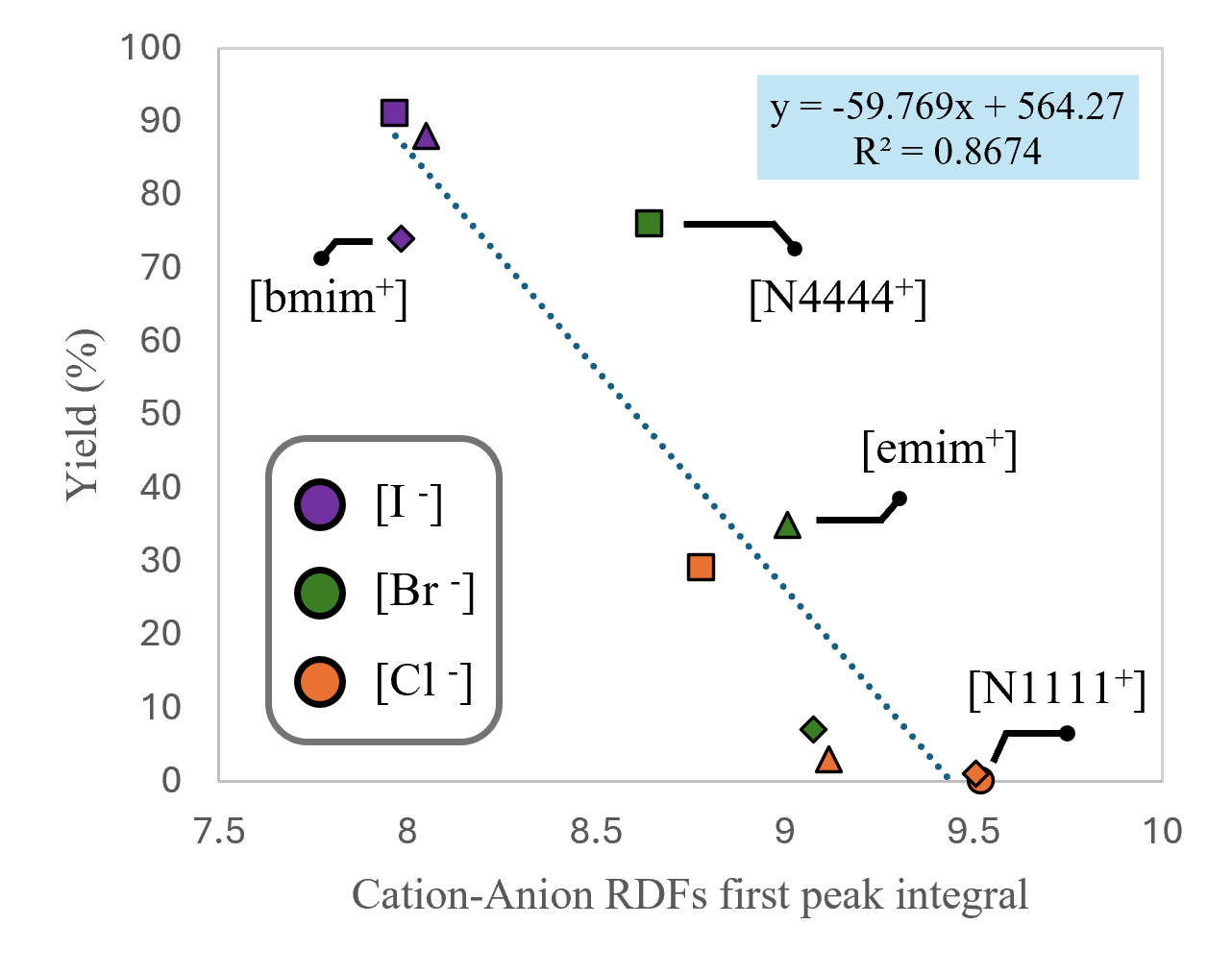}
    \caption{Experimental H-Epox yields (from \cite{hernandez2024dissociation}) against calculated area for each C-A RDF first peak of Fig.\ref{fig:RDFS-CA}.}
    \label{fig:Regress}
\end{figure}

\begin{figure}[h!]
    \centering
    \includegraphics[width=0.8\linewidth]{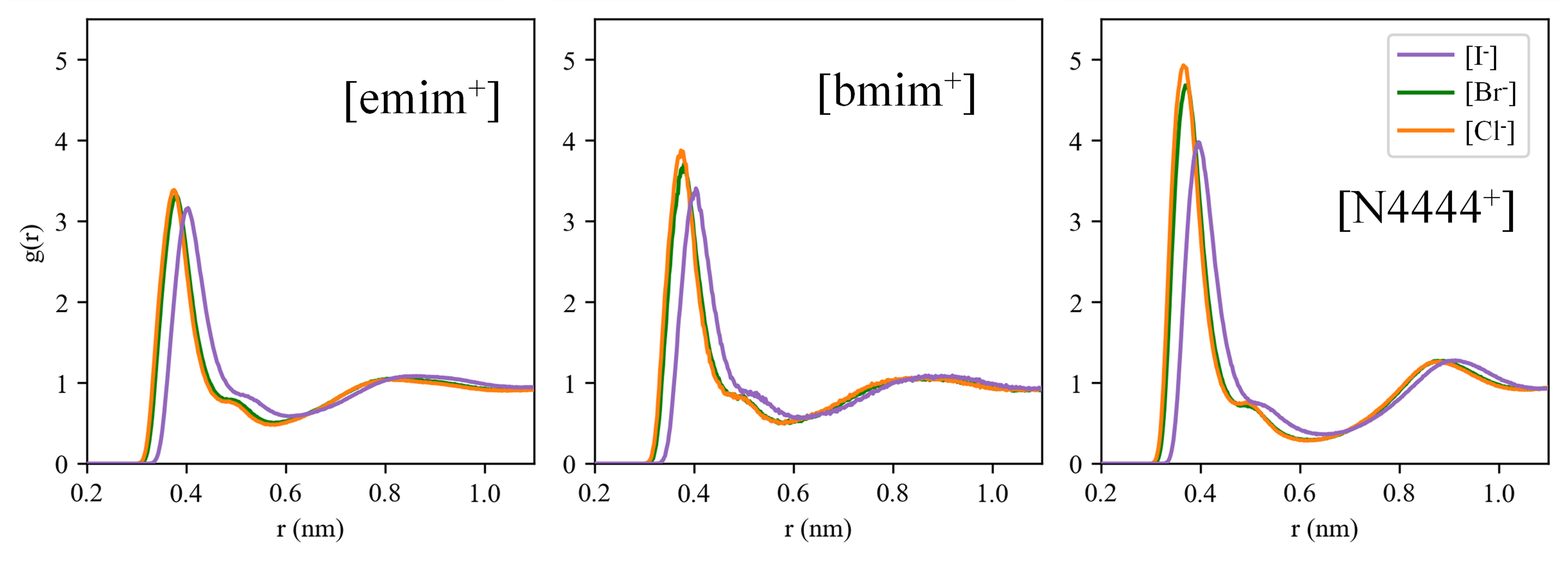}
    \caption{CH$_2$ epoxide carbon-anion (Epox-A) radial distribution functions (RDFs) obtained for each cation and halide anion combination.}
    \label{fig:RDFS-CarbA}
\end{figure}

\begin{figure}[h!]
    \centering
    \includegraphics[width=0.8\linewidth]{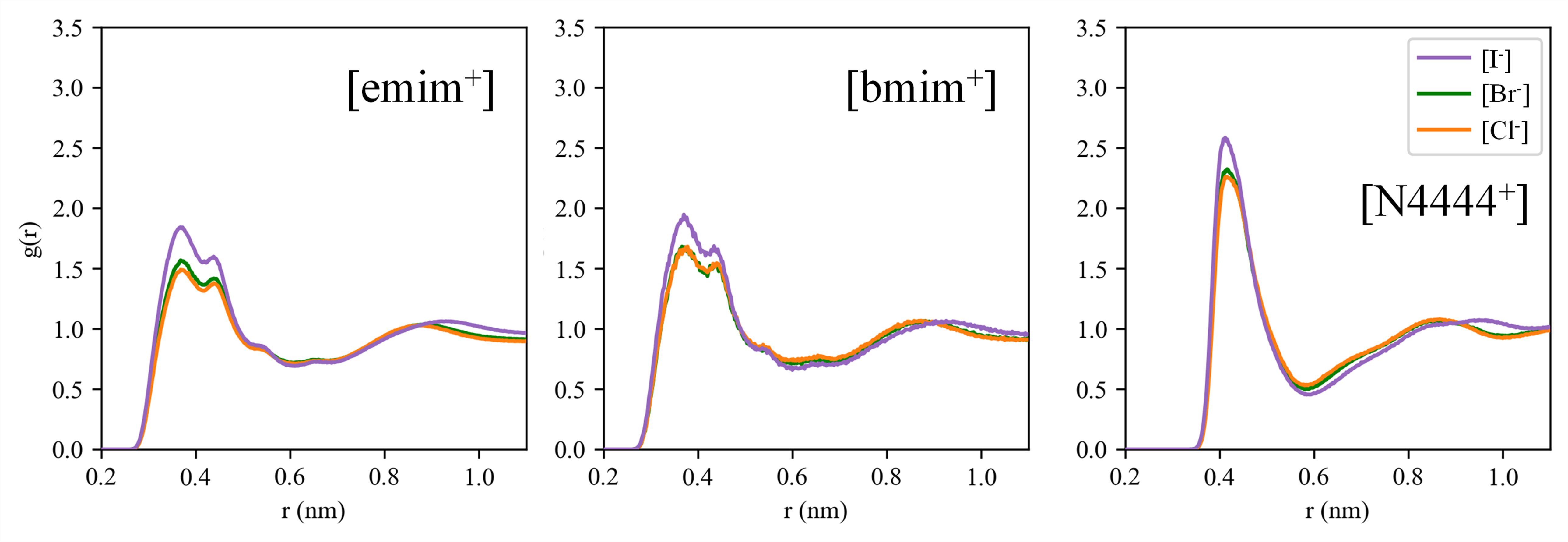}
    \caption{Cation-oxygen (C-Epox) radial distribution functions (RDFs) obtained for each cation and halide anion combination.}
    \label{fig:RDFS-CO}
\end{figure}

It should be noted that C-A RDFs (Fig.\ref{fig:RDFS-CA}) first peaks are much higher compared to those obtained for Epox-A (Fig. \ref{fig:RDFS-CarbA}) and C-Epox (Fig. \ref{fig:RDFS-CO}). Therefore, cation-anion interaction is by far the most probable interaction, meaning that in these ionic liquid systems, ions form ionic pairs or aggregates. The narrow peaks observed for [$\mathrm{N1111^+}$][$\mathrm{Br^-}$] IL (Fig. \ref{fig:N1111Br} in appendix) and [$\mathrm{N4444^+}$]-based ILs in Fig. \ref{fig:RDFS-CA}  reveal a predominant short cation-anion distance between the ionic species in reaction media, corresponding to the ion pairs observed previously in Figs. \ref{fig:N2222-solu} and  \ref{fig:N4444-solu}. Wider peaks are observed for [$\mathrm{emim^+}$] and [$\mathrm{bmim^+}$] cations, which correctly corresponds to the more dispersed charge distribution between the hydrogen atoms situated close to both heterocyclic nitrogens, which was also observed in Fig. \ref{fig:SDFS}. Additionally, for C-A (Fig. \ref{fig:RDFS-CA}) RDFs, the first peaks are higher for [$\mathrm{Br^-}$] and [$\mathrm{Cl^-}$] than for [$\mathrm{I^-}$], therefore having a greater area under the curve. This means that [$\mathrm{Br^-}$] and [$\mathrm{Cl^-}$] anions are slightly more associated with the cations. This result is promising to explain the experimental catalytic activity trend, as the most experimentally reactive halide, which corresponds to [$\mathrm{I^-}$] (\cite{hernandez2024dissociation}), is also the one that was obtained to be less associated with the cation in the simulations, meaning greater mobility and therefore a better availability for reaction. This result constitutes another validation of the force fields, since the conductivity measurements and dissociation constants follow this trend ([$\mathrm{I^-}$]$>$[$\mathrm{Br^-}$]$>$[$\mathrm{Cl^-}$])  \cite{hernandez2024dissociation}. This means that C-A RDFs are in great correspondence with the experimental evidence, therefore elucidating a possible mechanism underlying the catalytic activity. To test this, for each of the C-A RDFs, Fig. \ref{fig:RDFS-CA} (and Fig. \ref{fig:N1111Br}a from the appendix), the coordination number of the first peak was calculated for each system and correlated with the experimental yields (\cite{hernandez2024dissociation}), showing that a general correlation could not be established due to an insensitiveness of the obtained coordination numbers for a common cation, therefore neglecting the halide-anion effect (see Fig. \ref{fig:Coord_Num_Regress} in the appendix). As an alternative methodology, for each of the C-A RDFs, the area under the first peak (cumulative probability) multiplied by 10 was calculated by integration. By representing the experimental yields obtained from the literature \cite{hernandez2024dissociation} against each calculated area, Fig. \ref{fig:Regress} is obtained. It is worth noting that the regression coefficient obtained for the linear regression in the represented data results to be similar ($\approx$ 0.86) to the successful dissociation constant methodology recently demonstrated in \cite{hernandez2024dissociation}. These results enable to propose the methodology followed in this work as a possible pure computational methodology to estimate both solubility and yields of CO$_2$ cycloaddition to epoxides by halide-based ionic liquids, complementary to the K$_{D}$ computational descriptor provided by DFT/COSMO-RS approach.

By applying the same criteria, Epox-A RDFs in Fig.\ref{fig:RDFS-CarbA} find that [$\mathrm{Br^-}$] and [$\mathrm{Cl^-}$] are more likely to be found around the CH$_2$ carbon of the epoxide. So even though C-A interaction finds that [$\mathrm{Br^-}$] and [$\mathrm{Cl^-}$] are more attached to the cation and therefore should be less available for the reaction, it is additionally found that both are situated generally closer to the carbon susceptible to the nucleophilic attack. Two possibilities arise from this situation. The first is that, given the success description found in terms of the C-A RDFs, even though these anions are more capable of performing the nucleophilic attack to the epoxide, the intermediate formation (Int I in Fig.\ref{fig:mecanismo}) in which the anion is covalently bonded to the epoxide is impeded because of the strong interaction with the cation. The second possibility is the result obtained in the C-Epox RDFs set (Fig. \ref{fig:RDFS-CO}). This result shows that in [$\mathrm{I^-}$] simulations, the cation is located with greater probability nearer the epoxide oxygen (O), possibly promoting the activation of the epoxide by an electrophilic attack. This phenomenon, which as mentioned in the introduction, has been widely discussed in the bibliography \cite{bobbink2018intricacies, sun2007density, cheng2015effect, anthofer2014cycloaddition, yan2021fixation, clark2023solvation} and is consistent with the strong cation effect on the experimental cyclic carbonate yield \cite{hernandez2024dissociation}, is also confirmed in this work. This vision aligns with the reaction mechanism that proposes the epoxide activation as the first step, promoting afterward the nucleophilic attack of the anion. Both explanations seem rather plausible and in fact are not mutually exclusive. If cation-anion interaction is reduced and the anions try to situate near the CH$_2$ carbon, the cation will be able to interact better with the oxygen because of its better mobility than in a situation in which it is nailed to the anion. It is therefore thought that the reaction mechanism in which epoxide activation by the cation interaction with the oxygen, followed by the nucleophilic attack of the halide anion, is the most probable to be occurring in the studied systems, in good agreement with previous computational-experimental results \cite{yan2021fixation, bobbink2018intricacies}.

\section{Conclusions}
\label{sec:Conclusions}
In this work, molecular dynamics simulations were used aiming to describe the CO$_{2}$ cycloaddition to epoxides reaction using ionic liquids. Models set up, solubilities evaluation, and cation-anion interaction were inspected. \\

Relative to experimental or Peng-Robinson data, propylene oxide, and carbonate molecular dynamics simulations slightly underestimate vapor densities, overestimate liquid densities by $150\,\text{kg/m}^3$, and overestimate critical points. Despite these discrepancies, the force fields qualitatively capture density trends. The modifications made to the propylene oxide force field in combination with butyl cyclopropane to create the 1,2-epoxyhexane force field show a significant impact, over predicted liquid densities, being generally closer to Peng-Robinson predictions. Gas densities are underestimated at high temperatures (500 and 600 K), and liquid densities are underestimated at low temperatures (200 and 300 K) and overestimated at high temperatures (500 and 600 K). The critical point is overestimated compared to Peng-Robinson’s prediction around 600 K. The 300-500 K range best reproduces the densities, aligning with the experimental conditions of the experimental catalytic reference tests. The qualitative correctness of the model supports its continued use and the applicability of the methodology employed to hexylene carbonate.  \\

Ammonium-based ionic liquid's solubility in 1,2-epoxyhexane and hexylene carbonate is correctly reproduced with the CL\&P force field model in terms of crystallization and diffusion capability. [$\mathrm{N4444^+}$] cation-based ILs combined with any of the halide anions: [$\mathrm{I^-}$], [$\mathrm{Br^-}$] and [$\mathrm{Cl^-}$] result to be soluble in the carbonate/epoxide medium after long simulation times. [$\mathrm{N2222^+}$] cation-based ILs in combination with the halides resulted in the crystallization of the ionic liquids as independent pure phases, therefore being insoluble. Diffused cations and anions are closely interacting and in a similar proportion. This has led to propose that cations define solubility, which aligns with previous findings in the literature, and that the strong cation-anion interactions result in the formation of ion pairs or aggregates rather than dissociated free ions through the reactive media. \\

For [$\mathrm{N4444^+}$], [$\mathrm{emim^+}$], [$\mathrm{bmim^+}$] cations combined independently with the three halide anions [$\mathrm{I^-}$], [$\mathrm{Br^-}$] and [$\mathrm{Cl^-}$], and for [$\mathrm{N1111^+}$][$\mathrm{Br^-}$] IL, cation-anion interaction is the most predominant in the reaction media rather than ions interaction with 1,2-epoxyhexane atoms. Cation-anion interaction was quantified by the integration of the first peak of the computed radial distribution functions between the reference atoms in the cations and the halides. A correlation between the obtained areas and the experimental yields with a regression coefficient of 0.86 was established, which leads to propose that greater cation-anion interaction implies a reduction of the ionic liquid catalytic activity, aligning with previous findings in the literature, with the improvement of solubility prediction. Additionally,  ionic liquids composed of cations paired with [$\mathrm{I^-}$] anion, displayed an improved interaction between the cation and 1,2-epoxyhexane's oxygen compared to [$\mathrm{Br^-}$] and [$\mathrm{Cl^-}$], which is hypothesized to enable epoxide activation and reaction intermediates stabilization, which could explain the improved experimental yields displayed by these ionic liquids.

\section*{CRediT authorship contribution statement}

\textbf{Sergio Dorado-Alfaro}: Investigation, Writing - original draft, Writing – review \& editing. \textbf{Elisa Hernández}: Investigation, Writing – review \& editing. \textbf{Jesús Algaba}: Methodology, Supervision, Writing – review \& editing. \textbf{Pablo Navarro}: Conceptualization, Methodology, Supervision, Writing – review \& editing. \textbf{Felipe J. Blas}: Methodology, Supervision, Writing – review \& editing. \textbf{José Palomar}: Conceptualization, Investigation, Methodology, Supervision, Project administration, Writing – review \& editing.

\section*{Declaration of competing interest}

The authors declare that they have no known competing financial interests or personal relationships that could have appeared to influence the work reported in this paper.

\section*{Acknowledgements}

The authors are grateful to Ministerio de Ciencia e Innovación of
Spain (projects TED2021-129803A-I00 and PID2023-150532OB-I00)
for financial support and Centro de Computación Científica de la Universidad Autónoma de Madrid for computational facilities. SDA aknowledges the FPI Grant (PREP2023-001656) from Ministerio de Ciencia e Innovación, Agencia Estatal de Investigación and Fondo Social Europeo Plus. JA and FJB also acknowledge funding from Ministerio de Ciencia e Innovaci\'on (Grant No.~PID2021-125081NB-I00) and Universidad de Huelva (P.O.~FEDER~EPIT1282023), both co-financed by EU FEDER funds. We also thank the Centro de Su- percomputación de Galicia (CESGA, www.cesga.es) for providing acess to the computing resources of the Finisterrae III Supercomputer.

\section*{Data availability}

No data was used for the research described in the article.

 \bibliographystyle{elsarticle-num} 
 \bibliography{cas-refs}






\appendix

\section{Appendix Section}
\label{sec:sample:appendix}

\begin{figure}[h!]
    \centering
    \includegraphics[width=0.7\linewidth]{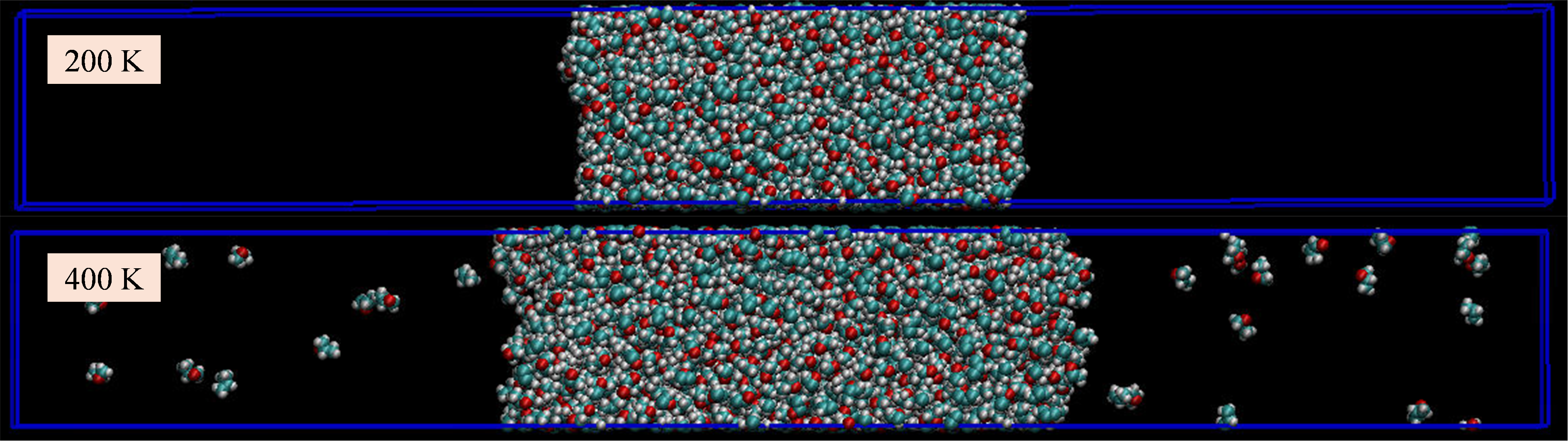}
    \caption{Visual VMD representation of propylene oxide (P-Epox) $NVT$ simulation boxes at 200 and 400 K as obtained from molecular dynamic simulations.}
    \label{fig:Visual_P-Epox}
\end{figure}

\begin{figure}[h!]
    \centering
    \includegraphics[width=\linewidth]{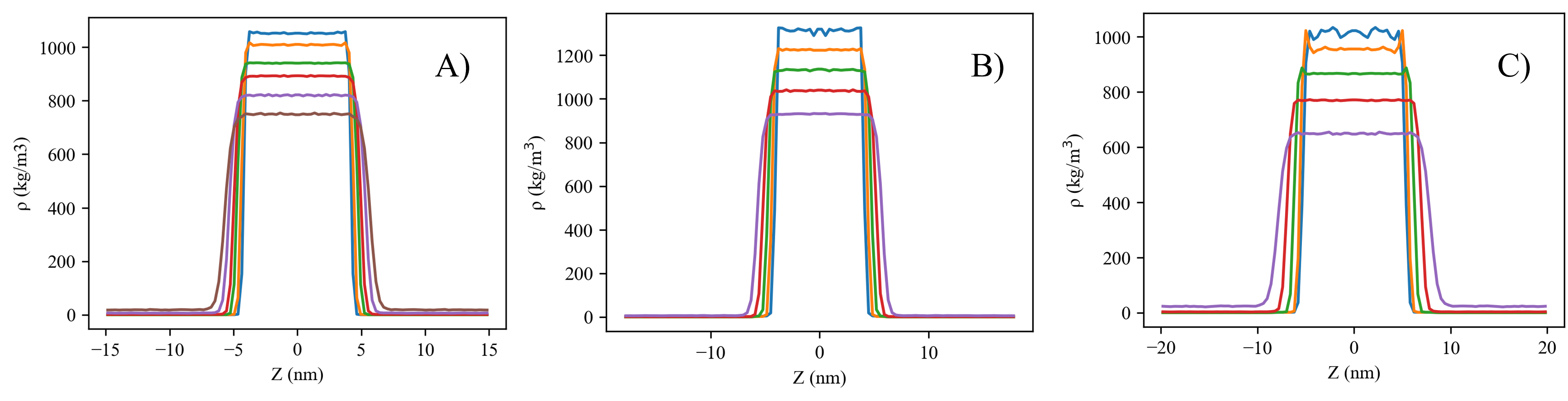}
    \caption{Density profile results for A) P-Epox, B) P-Carb, and C) H-Epox Vapor-Liquid equilibrium as obtained from molecular dynamic simulations.}
    \label{fig:1SUB}
\end{figure}

\begin{figure}[h!]
    \centering
    \includegraphics[width=\linewidth]{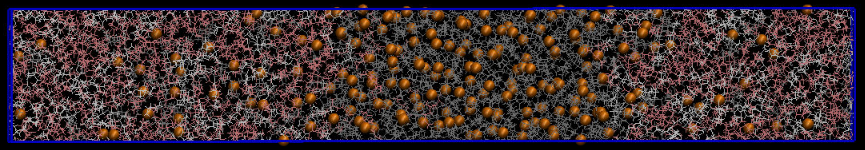}
    \caption{Visual VMD representation of [$\mathrm{N4444^+}$][$\mathrm{Cl^-}$], H-Epox (pink) and H-Carb (white) obtained from semi-isotropic $NPT$ simulation. H-Epox and H-Carb models result mutually soluble from the initial independent pure phases.}
    \label{fig:todos-solu}
\end{figure}

\begin{figure}[h!]
    \centering
    \includegraphics[width=\linewidth]{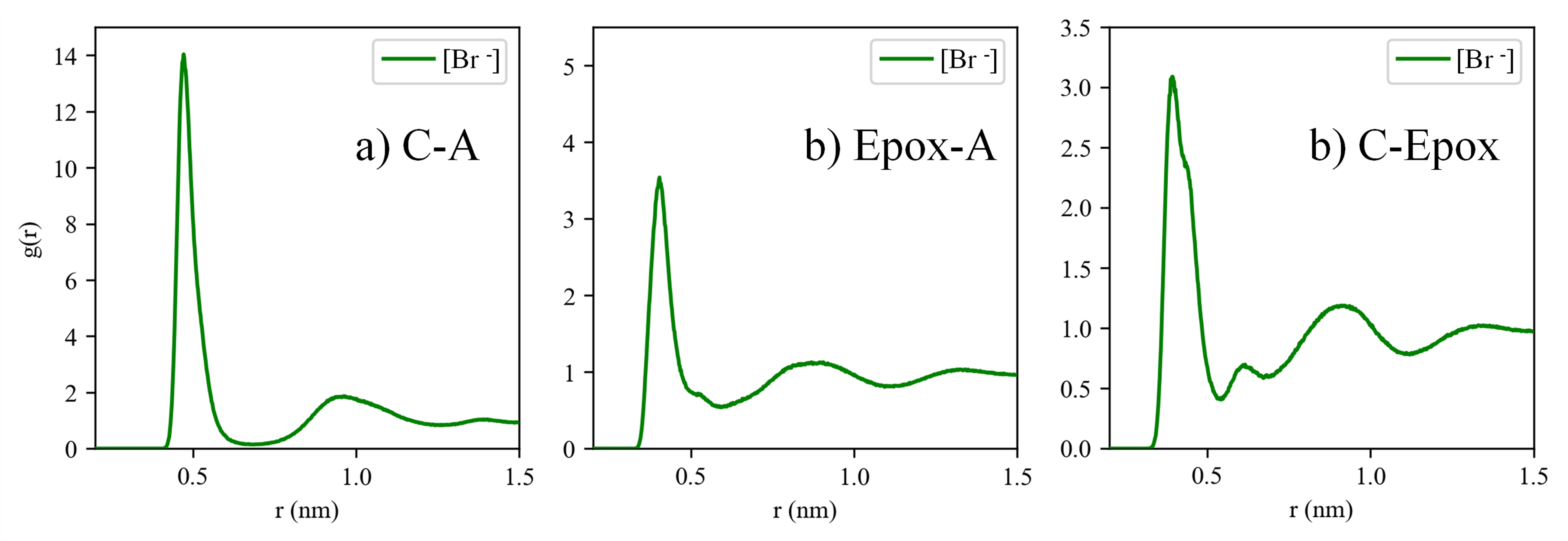}
    \caption{a) Cation-Anion (C-A) b) CH$_2$ epoxide carbon-Anion (Epox-A) and c) Cation-oxygen (C-Epox) radial distribution functions (RDFs) obtained for [$\mathrm{N1111^+}$][$\mathrm{Br^-}$] ionic liquid.}
    \label{fig:N1111Br}
\end{figure}

\begin{figure}[h!]
    \centering
    \includegraphics[width=0.7\linewidth]{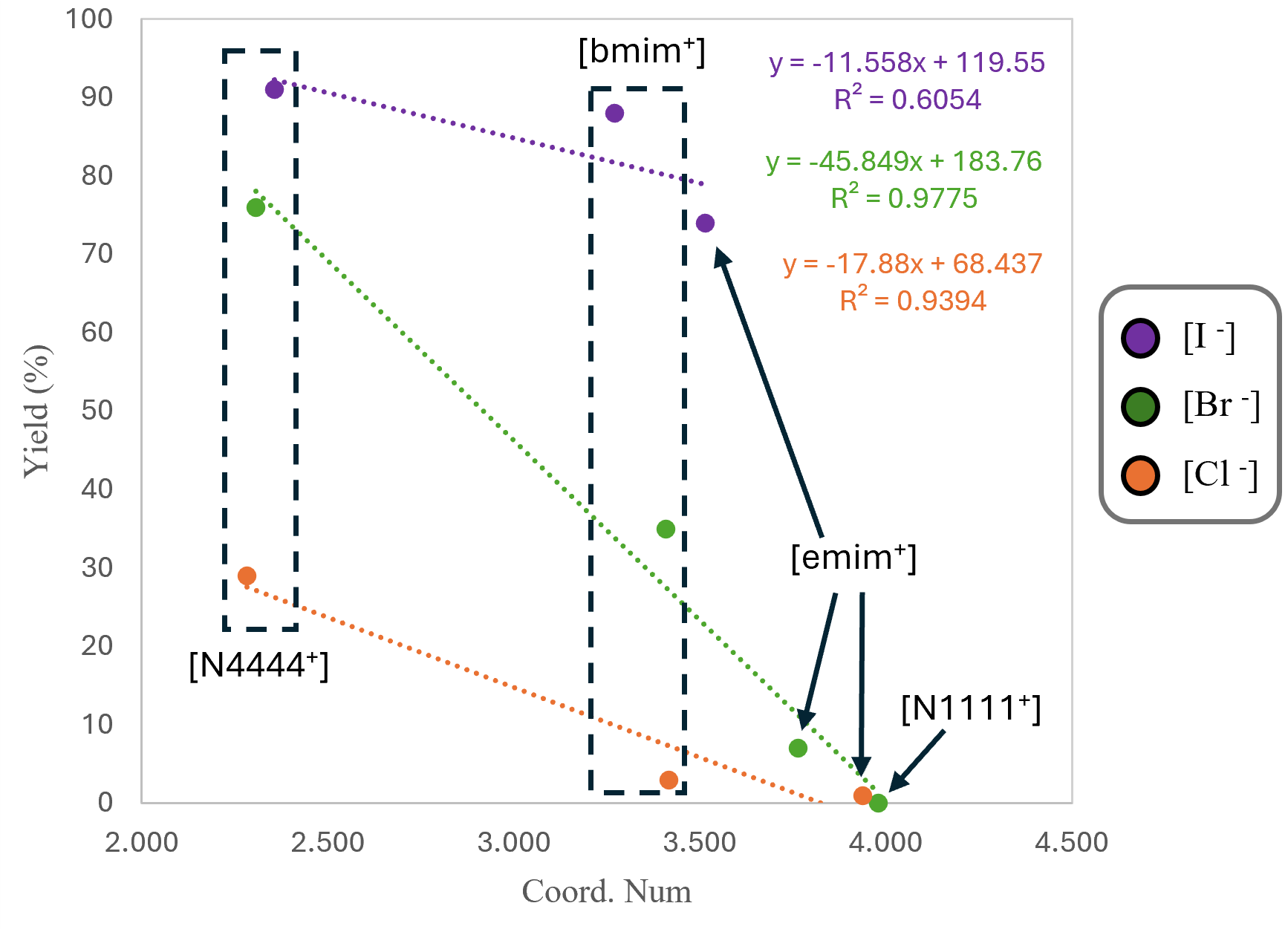}
    \caption{Experimental H-Epox yields (from \cite{hernandez2024dissociation}) against the first peak coordination number for each C-A RDF of Fig.\ref{fig:RDFS-CA}.}
    \label{fig:Coord_Num_Regress}
\end{figure}

\end{document}